\documentclass[amsmath,amssymb,aps,pra, onecolumn,notitlepage,superscriptaddress]{revtex4-1}

\usepackage{graphicx}% Include figure files
\usepackage{dcolumn}% Align table columns on decimal point
\usepackage{bm}% bold math
\usepackage{subcaption}
\usepackage{lettrine}
\usepackage{color}
\definecolor{green}{rgb}{0,0.6,0}
\definecolor{blue}{rgb}{0,0,0.6}
\definecolor{orange}{rgb}{0.93,0.69,0.13}
\usepackage{siunitx}
\DeclareSIUnit\gauss{G}

\begin{document}

\title{\bf Long-range vs. short-range effects in cold molecular ion-neutral collisions: \\ Charge exchange of Rb with N$^+_2$ and O$^+_2$}
%\author{Alexander D. D\"orfler$^1$, Pascal Eberle$^1$, Debasish Koner$^1$, \\ Micha\l{} Tomza$^{2\dagger}$, Markus Meuwly$^{1\ast}$ and Stefan Willitsch$^{1\ast}$}

\author{Alexander D. D\"orfler, Pascal Eberle, Debasish Koner}
\affiliation{Department of Chemistry, University of Basel, Klingelbergstrasse 80, 4056 Basel, Switzerland}
\author{Micha\l{} Tomza}\email{michal.tomza@fuw.edu.pl}
\affiliation{Faculty of Physics, University of Warsaw, Pasteura 5, 02-093 Warsaw, Poland}

\author{Markus Meuwly}\email{m.meuwly@unibas.ch}
\affiliation{Department of Chemistry, University of Basel, Klingelbergstrasse 80, 4056 Basel, Switzerland}

\author{Stefan Willitsch}\email{stefan.willitsch@unibas.ch}
\affiliation{Department of Chemistry, University of Basel, Klingelbergstrasse 80, 4056 Basel, Switzerland}


\begin{abstract}

\noindent We report a study of cold charge-transfer (CT) collisions of Rb atoms with N$_2^+$ and O$_2^+$ ions in the mK regime using a dynamic ion-neutral hybrid trapping experiment. State- and collision-energy-dependent reaction rate coefficients have been measured for both systems. We observe markedly different charge-transfer kinetics and dynamics for the different systems and reaction channels. While the kinetics in some channels are consistent with classical capture theory for the rate coefficient, others show distinct non-universal dynamics. The experimental results are interpreted with the help of classical capture, quasiclassical trajectory and quantum scattering calculations using state-of-the-art ab-initio potentials for the highly excited molecular states involved. The theoretical analysis reveals an intricate interplay between short- and long-range effects in the different reaction channels which ultimately determines the CT dynamics and rates. At short range, CT was found to occur via both single and multiple collision events with the latter typically showing pronounced large-amplitude internal motions of the collision complex. Our results illustrate salient mechanisms that determine the efficiency of cold molecular CT reactions.
\end{abstract}
\maketitle

\section{Introduction}
\label{sec: introduction}
\noindent Studies of ion-neutral interactions at very low temperatures have progressed considerably in recent years as a result of the development of techniques for the combined ("hybrid") trapping of cold atoms and ions~\cite{willitsch12a, willitsch15a, willitsch17a, sias14a,haerter14a,tomza17a}. At temperatures in the mK regime achievable in these experiments, new possibilities open up for the detailed exploration of collisional and chemical dynamics at the quantum level. Interactions between cold atomic ions and neutral atoms have been widely studied with these setups over the last decade. These investigations have provided new insights into cold reactive processes~\citep{hall11a,rellergert11a,schmid10a,hall13b,joger17a}, the sympathetic and internal cooling of ions by ultracold atoms~\cite{zipkes10a,rellergert13a,haze17a}, cold three-body recombination dynamics~\cite{krukow16a}, spin-exchange and -relaxation processes~\citep{sikorsky18} and the statistical mechanics of trapped ions in a cold buffer gas~\citep{meir16b,rouse17a}. 
\noindent While a large body of data by now exists on a range of atomic collisions systems, studies of cold collisions with molecular ions are still sparse. Previous experiments included the combination of cold molecular beams or cryogenic gases with trapped ions~\citep{willitsch08a, otto08a}, the merging of molecular beams containing Rydberg molecules~\cite{allmendinger16a} and the sympathetic cooling of molecular ions in hybrid trapping experiments~\citep{hall12a,rellergert13a,puri17a,puri19a}. These latter studies uncovered unusually fast kinetics~\cite{hall12a}, the formation of exotic molecular species~\cite{puri17a} and reaction blockading of short-lived excited species \cite{puri19a} in cold molecular ion-neutral atom systems. These investigations gave a first glimpse at the wealth of additional phenomena which can be explored by extending hybrid trapping experiments from atomic to molecular systems.
\noindent In most of the systems studied so far in hybrid trapping experiments, charge transfer (CT) between the neutral atoms and ions was found to be a dominating reactive process. CT can be promoted either by radiative coupling of the entrance channel to energetically lower-lying CT channels~\cite{dasilva15a} or by non-adiabatic coupling between channels~\cite{tacconi11a}. In a previous study on reactions between N$_2^+$ molecular ions and Rb atoms~\cite{hall12a} at collision energies around 20\,mK, a strong dependence of the CT rate coefficient on the electronic state of Rb was observed. CT was found to be extremely fast in collisions with excited Rb~$(5p)~^2P_{3/2}$ atoms. By contrast, reactions with ground-state Rb~$(5s)~^2S_{1/2}$ atoms were observed to be considerably slower than the Langevin limit which frequently serves as a benchmark for ion-molecule reactions~\cite{zhang17a}. In the Langevin picture, the rate coefficient is limited by capture of the collision partners by long-range ion-induced dipole interactions whereas the short-range reaction probability is assumed to be unity~\cite{gioumousis58a}.
The results of Ref.~\cite{hall12a} already hinted at the importance of the interplay between short- and long-range effects in cold CT reactions. Motivated by these findings, we explore this topic here in a comparative study of the CT between molecular oxygen ($^{32}$O$_2^+$) and nitrogen ($^{28}$N$_2^+$) ions with $^{87}$Rb atoms using experiments in the cold regime together with computational work to interpret the dynamics at a molecular level. 
\begin{figure}[htbp!]
	\centering
		\includegraphics[width=0.95\textwidth]{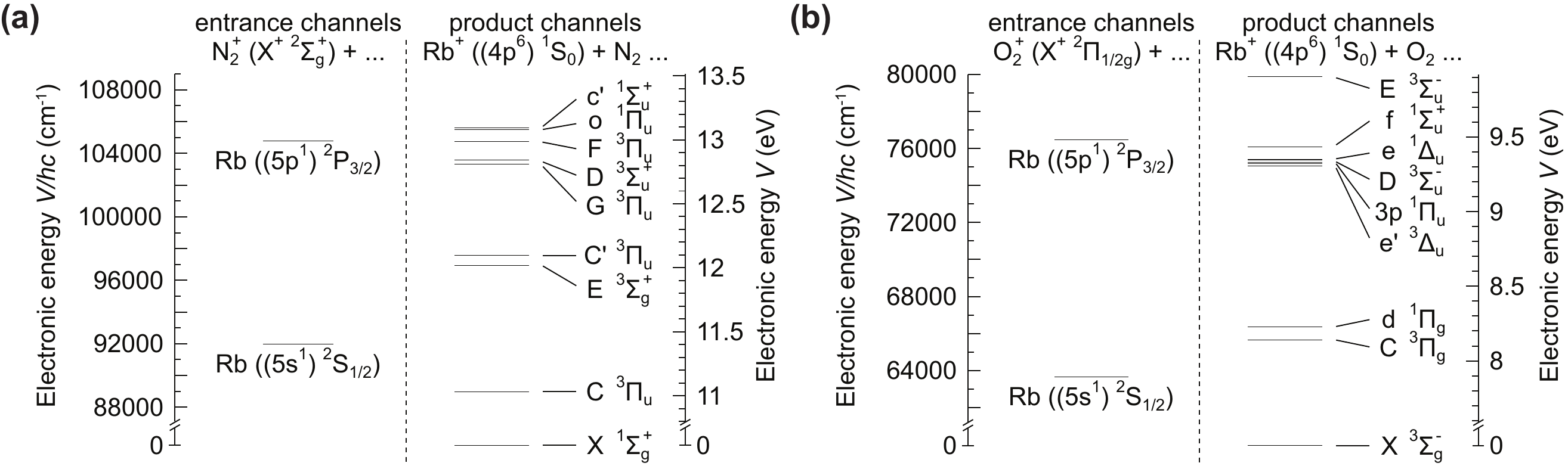}%{energylevels}
 	\caption{Asymptotic energies of the entrance and near-resonant product channels of (a) N$^+_2$+Rb and (b) O$^+_2$+Rb charge-transfer (CT) collisions. The molecular ions can undergo CT with Rb in either its $(5s)~^2S_{1/2}$ ground or $(5p)~^2P_{3/2}$ excited state populated by laser cooling in a magneto-optical trap. All energies are referenced to the asymptotes of the lowest product channels connecting to the ground state of the relevant neutral molecules. See text for details.}
	\label{fig:elvl}
\end{figure}
\noindent The energetics of the reactions considered are illustrated in Fig.~\ref{fig:elvl}. N$_2^+$ and O$_2^+$ ions in their electronic ground states collide with Rb atoms in either the $(5s)~^2S_{1/2}$ ground or $(5p)~^2P_{3/2}$ first excited electronic state to produce Rb$^+$ ions in the $^1S_{0}$ ground electronic state and neutral N$_2$\,/\,O$_2$ molecules. On purely energetic grounds, the molecular CT products can form in a range of highly exited electronic states in both cases, in particular in collisions with excited Rb atoms in which case a range of low Rydberg states of the neutral product molecules are energetically accessible. Fig.~\ref{fig:elvl} also makes apparent that the entrance channels of the reactions considered here correspond to highly excited electronic states of the collision system.
\noindent In a first approximation, CT is considered to be most efficient if it is near resonant, i.e., if the entrance and product channels are energetically similar, and if it does not involve a marked reconfiguration of the electrons in the molecule~\cite{vanderkamp94a}. On these grounds, it could be expected that CT with Rb atoms in the $^2P_{3/2}$ excited state should be more efficient owing to the higher density of near-resonant, electronically favourable product channels (see Fig.~\ref{fig:elvl}).
\noindent Employing recently established experimental methods which allow an improved control over both the electronic state of the reactants as well as the collision energy in the mK regime~\cite{eberle16a}, we studied the kinetics of CT in the different collision channels. We found a marked dependence of the dynamics on the initial state of Rb. CT was observed to be generally fast and the trends were found in many cases not to be compatible with the expectations outlined above. The experiments were analysed and modelled with the help of electronic-structure as well as quasiclassical trajectory simulations and quantum-dynamics calculations. This theoretical modelling revealed an intricate interplay of long-range interactions and localised short-range nonadiabatic couplings which, guided by the topology of the relevant potential-energy surfaces, determine the details of the CT dynamics and kinetics. This situation stands in stark contrast to cold CT collisions in atomic systems which are usually slow and often dominated by radiative couplings~\cite{willitsch15a,sias14a,haerter14a,tomza17a}.
\section{Experimental and Theoretical Methods}
\label{sec:methods}
\paragraph{Experimental methods.}
The experimental setup used in the present study has been described in detail previously~\cite{hall11a, hall13a, eberle16a}. Briefly, an ion-neutral hybrid trap was implemented by superposing a linear rf trap~\citep{willitsch08a} for the trapping and cooling of ions with a magneto-optical trap (MOT) for $^{87}$Rb atoms~\citep{hall11a}. The ion trap was operated at a frequency of \SI{3.25}{\mega\hertz} with an amplitude of $V_{\text{rf}}=\SI{400}{\volt}$ and featured 12 separately addressable electrodes for applying static and rf voltages. An atomic beam of Ca was generated from a resistively heated oven from which Ca$^+$ ions were loaded into the trap by non-resonant photoionization. The Ca$^+$ ions were subsequently laser cooled to form Coulomb crystals~\citep{willitsch12a}. Molecular ions were generated inside the trap from photoionization of room temperature background gas at a background pressure of \SI{1e-8}{\milli\bar} using a $\left[2+1\right]$ resonance-enhanced multiphoton ionization (REMPI) via the $a^{''}~^1\Sigma^+_g$ electronic state for N$_2$~\cite{tong10a} and via the $^3\Phi_g$($\nu=1$) Rydberg state for O$_2$~\cite{wang93b}. The molecular ions were sympathetically cooled by the Ca$^+$ ions to form strings localised on the rf null line of the trap. Following photoionization, the background pressure was kept at \SI{1e-8}{\milli\bar} for 30\,seconds to allow collisions to establish a room temperature distribution of rotational-state populations in the vibrational ground state of the ions. An EMCCD camera coupled to a microscope was used to obtain images of the Coulomb crystals by collection of the spatially resolved fluorescence of the trapped ions. The MOT was continuously loaded from background Rb vapor replenished by an alkali-metal dispenser.  The MOT is capable of operating in three modes: stationary operation, bright shuttling and dark shuttling. In the stationary mode, the cold atom cloud was superimposed on the ions, while in shuttling mode the cold atoms were repeatedly shuttled through the ions at well defined velocities using radiation pressure forces~\cite{eberle16a} enabling the tuning of the collision energies in the experiments. In the bright shuttling mode, the transversal Rb cooling lasers were left on so that parts of the Rb atoms were excited to the $^2P_{3/2}$ state during transit. In the dark shuttling mode, all lasers were switched off after accelerating the atom cloud so that the populations were confined to the ground state. Thus, while the static mode enables the measurement of CT rate coefficients as a function of the Rb state populations and, therefore, the determination of state-dependent rate coefficients, the dynamic mode allows in addition the measurement of reaction rates as a function of the collision energy with the Rb atoms in either the ground or a mixture of the ground and excited states. 
\paragraph{Electronic structure calculations.}
The potential energy surfaces for the ground and excited electronic states were obtained using the multireference configuration interaction method restricted to single and double excitations, MRCISD, starting from orbitals obtained with the multi-configurational self-consistent field method, MCSCF~\cite{WernerJCP88}. All atomic valence orbitals, i.e. both binding and antibinding molecular orbitals, were included in the complete-active-space (CAS) reference wave functions. The N and O atoms were described using the augmented correlation-consistent polarised core-valence quintuple-$\zeta$ quality basis sets (aug-cc-pCV5Z)~\cite{DunningJCP89}. The scalar relativistic effects in Rb were included by employing the small-core relativistic energy-consistent pseudopotential ECP28MDF to replace the inner-shells electrons~\cite{LimJCP05}, while remaining electrons were described with the large $[14s14p7d6f1g]$ basis set~\cite{TomzaMP13}. The electronic structure calculations were performed with the \textsc{Molpro} package of \textit{ab initio} programs~\cite{molpro}. {\it Ab initio} energies were calculated for two excited $^3$A$'$ electronic states of RbN$_2^+$ on a two-dimensional grid in Jacobi coordinates ($R, \theta$) for a fixed N-N distance $r=2.074$\,a.u. corresponding to the equilibrium bond length in N$_2^+$. Here, $R$ is the distance from the centre of mass of N$_2$ to Rb, $r$ is the distance between the two N atoms and $\theta$ is the angle between $\vec{r}$ and $\vec{R}$. The upper PES adiabatically correlates with the N$_2^+$(X$^{'2}\Sigma_g^+$)\,+\,Rb($^2$S$_{1/2}$) asymptote while the lower surface dissociates towards N$_2$(C$^{3}\Pi_u$)+Rb$^+$. Two-dimensional analytical PESs for the two coupled $^3$A$'$ electronic states were constructed from the {\it ab initio} energies using the reproducing kernel Hilbert space (RKHS) technique~\cite{ho96:2584,unk17:1923}. For the radial dimension ($R$), a reciprocal power decay kernel was used which smoothly decays to zero $\propto \frac{1}{R^4}$ and gives the correct long-range behavior for ion-neutral type interactions. For the angular degree of freedom, a Taylor spline kernel was used.
\paragraph{Quantum scattering calculations.}
The rate coefficients for the non-adiabatic CT collisions were calculated along one-dimensional cuts of the two crossing potential energy surfaces in a diabatic representation and subsequently integrated over all possible orientation angles. The coupled-channels equations for the nuclear motions were solved using a renormalized Numerov propagator~\cite{JohnsonJCP78} with step-size doubling as implemented in Ref.~\cite{TomzaPRL14}. The ratios of the wave function at two adjacent grid points were propagated to large particle separations $R$, where the $K$ and $S$ matrices were extracted by imposing long-range scattering boundary conditions in terms of Bessel functions. The inelastic rate coefficients were obtained from the elements of the $S$ matrix summed over all relevant partial waves~$l$ and thermally averaged. 
\paragraph{Quasiclassical dynamics simulations.}
The quasiclassical-trajectory (QCT) method followed in this work has been discussed in detail in Ref.~\cite{kon18:094305} based on Refs.~\cite{tru79} and~\cite{hen11}. Hamilton's equations of motion were solved using a sixth-order symplectic method. Initial conditions for a trajectory were sampled from a standard Monte Carlo sampling method~\cite{tru79}. The rotational states of N$_2^+$ were sampled from a Boltzmann distribution at room temperature (300\,K). Stratified sampling~\cite{tru79, ben15:054304} was used to sample the impact parameter $b$. Two time steps ($\Delta t$) of 0.6 (from the beginning of a trajectory until it reaches $R < 35$\,a.u.) and of 0.05\,fs (for the first time a trajectory reaches $R < 35$\,a.u. to the end) were used for the numerical integration to ensure conservation of total energy and total angular momentum. The trajectory surface hopping (TSH) method~\cite{sti76:3975} was used to determine charge transfer rates including nonadiabatic transitions within a modified Landau-Zener~\cite{lan32:46,zen32:696,bel11:014701,bel14:224108} formalism. The modified Ladau-Zener formula depends on the adiabatic potential energies of the states involved in the transition and at time $t_c$ is,
\begin{equation}
 P_{\rm LZ}^{j\rightarrow k} = {\rm exp} \left( - \frac{\pi}{2\hbar}
 \sqrt{\frac{\Delta V^a_{jk}(x(t_c))^3}{\frac{d^2}{dt^2}\Delta
 V^a_{jk}(x(t_c))}} \ \right).
\end{equation}
Here, $P_{\rm LZ}^{j\rightarrow k}$ is the transition probability from state $j$ to state $k$ and $\Delta V^a_{jk}(x)$ is the adiabatic energy difference between these states. Whenever $\Delta V^a_{jk}(x)$ reached a local minimum, transition probabilities were calculated and momentum corrections along different degrees of freedom were applied~\cite{mil72:5637} after a successful hop to keep the total energy and angular momentum conserved for a given trajectory. The rate coefficients at a particular collision energy ($E_{\rm coll}$) were calculated from
\begin{equation}
 k = g_e \sqrt{\frac{2 E_{\rm t}}{\mu}} \pi b^2_{\rm max} \frac{N_{\rm
 r}}{N_{\rm tot}},
\end{equation} where $\mu$ is the reduced mass of the collision system, $N_{\rm r}$ is the number of reactive (charge transfer) trajectories (weighted by stratum statistical weight), $N_{\rm tot}$ is the total number of trajectories, $g_e$ is the electronic degeneracy factor (here 3/4)  and $b_{\rm max}$ is the maximum impact parameter for which a charge transfer reaction can occur. 
\section{Results and discussion}
\label{sec: results}
\subsection{State-dependent charge-transfer rate coefficients}
\label{subsec:popdep}
\begin{figure}[htbp!]
	\centering
		\includegraphics[width=1\columnwidth]{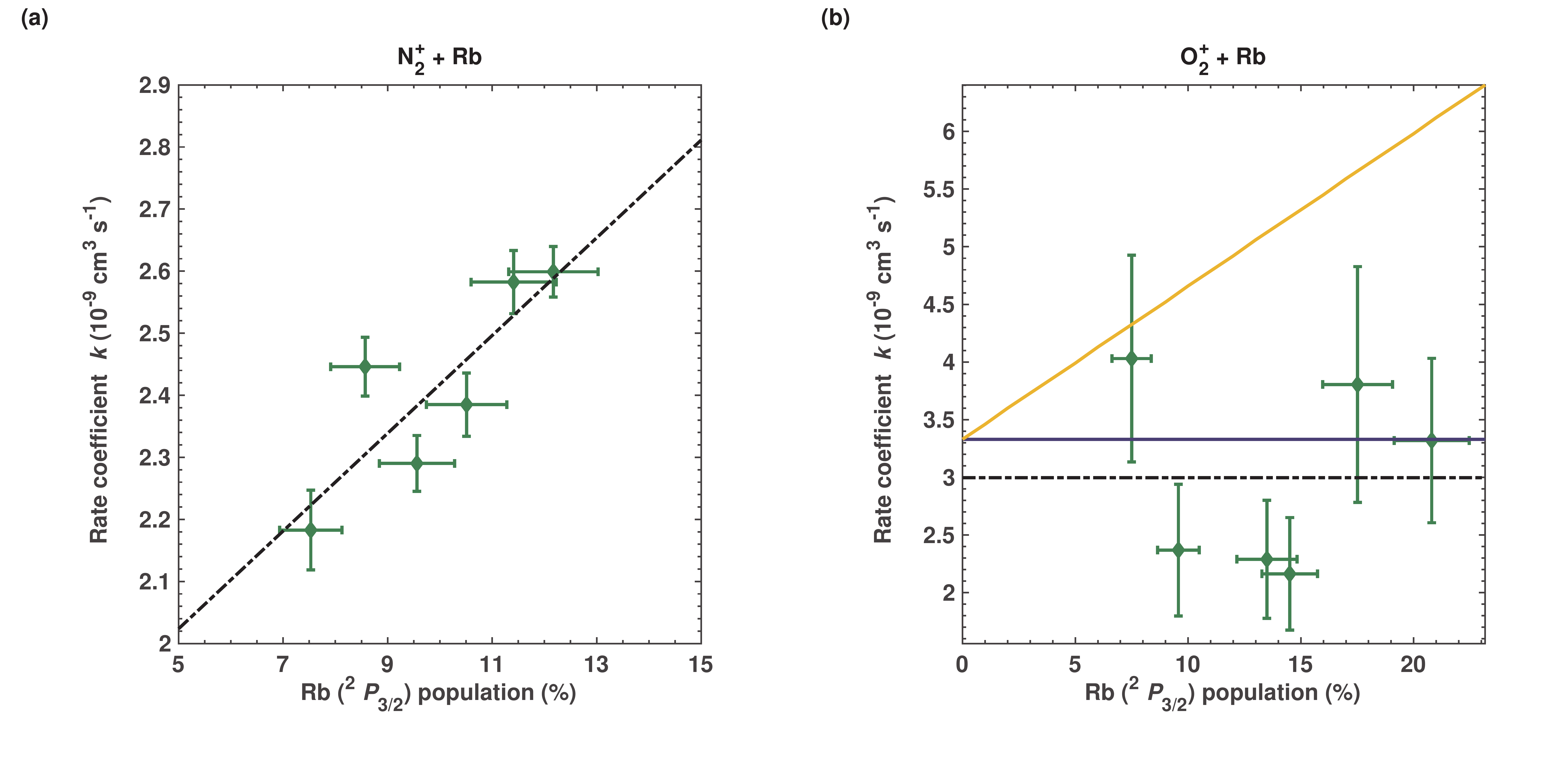}%{stationary MOT reactions}
 	\caption{Effective state-averaged CT reaction-rate coefficients as a function of Rb($^2P_{3/2}$) population for (a) N$^+_2$ and (b) O$^+_2$. The black dash-dotted lines show linear fits to the experimental data (green symbols). The orange solid line in (b) is a prediction of the effective rate coefficient for O$_2^+$\,+\,Rb assuming Langevin interactions in reactions with Rb~$^2S_{1/2}$ and additional ion-quadrupole interactions in reactions with Rb~$^2P_{3/2}$. The blue solid line indicates the theoretical Langevin rate coefficient for O$^+_2$\,+\,Rb($^2S_{1/2}$). See text for details. Error bars correspond to one standard error.}
	\label{fig:stat}
\end{figure}
\noindent Fig.~\ref{fig:stat} shows the dependence of the effective, i.e., state-averaged, CT rate coefficient as a function of the Rb($^2$P$_{3/2}$) population for reactions with N$^+_2$ in Fig.~\ref{fig:stat}~(a) and with O$^+_2$ in Fig.~\ref{fig:stat}~(b). The green symbols represent the experimental results and the dash-dotted lines are linear fits to the data. For N$_2^+$\,+\,Rb (Fig.~\ref{fig:stat}~(a)), a clear dependence of the rate coefficient on the Rb($^2P_{3/2}$) population is observed. These results can be compared with our previous study~\cite{hall12a} in which this observation has been rationalised by an enhanced CT rate in the excited channel dominated by the capture of the ion by the permanent quadrupole of Rb in the $^2P_{3/2}$ state~\cite{hall12a}. The fit yields a rate coefficient $k_p(\text{N}_2)$=\SI{1.7(6)e-8}{\cubic\centi\meter\per\second} for reactions with Rb($^2P_{3/2}$) (in line with the result of~\cite{hall12a}) and $k_s(\text{N}_2)$=\SI{1.6(3)e-9}{\cubic\centi\meter\per\second} for reactions with Rb($^2S_{1/2}$). This new value for $k_s$ replaces the previous upper-bound estimate of  $\leq$~\SI{2e-10}{\cubic\centi\meter\per\second} of Ref.~\citep{hall12a}. 
\noindent By contrast, the results for O$^+_2$\,+\,Rb shown in Fig.~\ref{fig:stat}~(b) reveal no clear dependence of the effective CT rate on the Rb excited-state population. The hypothetical case of an effective rate coefficient expected for Langevin capture in the ground channel and, as with N$_2^+$, dominant ion-quadrupole capture in the excited channel (solid orange line) does not agree with the observed data. Surprisingly, the effective rate coefficient closely follows a limiting value set by Langevin theory (blue solid line), implying Langevin-type dynamics in both the ground and excited channels of O$_2^+$\,+\,Rb. The Langevin rate coefficient for O$^+_2$+Rb($^2S_{1/2}$) is $k_s^\text{L}(\text{O}_2)$=\SI{3.3e-9}{\cubic\centi\meter\per\second} (blue solid line) and a fit of a constant function to the data yields an effective rate coefficient $k_\text{eff}(\text{O}_2)$=\SI{2.9(8)e-9}{\cubic\centi\meter\per\second} (black dash-dotted line).
\subsection{Collision-energy dependent charge-transfer rate coefficients}
\label{subsec:endep}
\begin{figure}[htbp!]
	\centering
		\includegraphics[width=1\columnwidth]{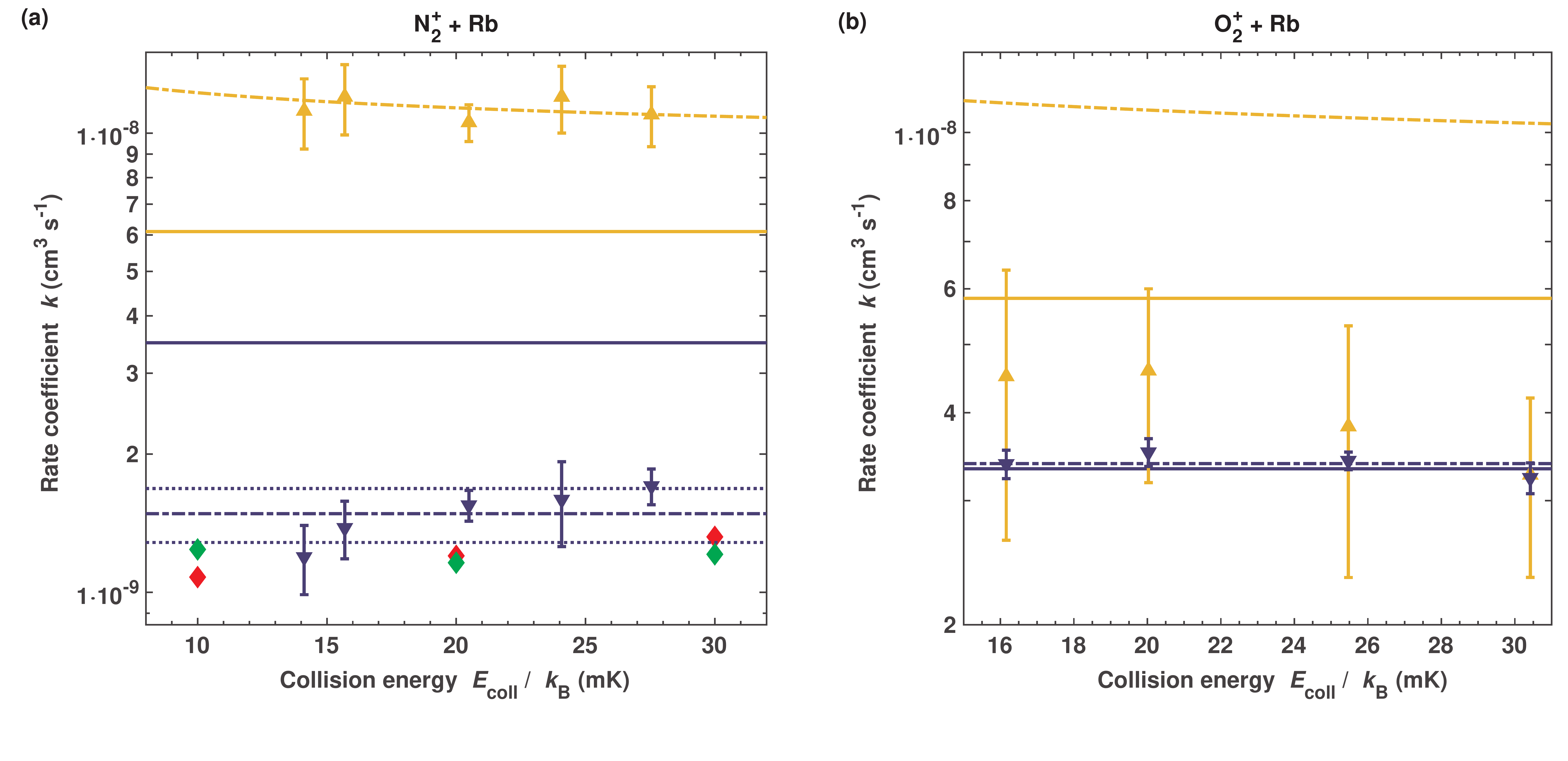}%{shuttling MOT reactions}
		\caption{Reaction rate coefficients as a function of collision energy and state of Rb for (a) N$^+_2$\,+\,Rb and (b) O$^+_2$\,+\,Rb. Blue [orange] symbols indicate CT rate coefficients for collisions with Rb in its $^2S_{1/2}$ [$^2P_{3/2}$] state. The theoretical Langevin rate coefficients for collisions in these channels are the blue [orange] solid lines. Dash-dotted lines represent fits to the experimental data, except for the orange dash-dotted line in (b) which represents the prediction of the rate coefficient for dominant ion-quadrupole capture. The red and green diamonds in (a) represent CT rate coefficients for the N$_2^+$\,+\,Rb $(^2S_{1/2})$ computed by quasiclassical trajectory simulations and quantum scattering calculations, respectively. See text for details. Error bars represent one standard error.}
	\label{fig:shut}
\end{figure}
\noindent The dynamic mode of the experiment allows both collision-energy- and state-dependent measurements of CT rate coefficients by enabling or disabling laser cooling of the Rb atoms during the shuttling process. Fig.~\ref{fig:shut} shows the collision-energy and state-dependent rate coefficients for (a) N$^+_2$+Rb and (b) O$^+_2$+Rb in the collision-energy range from $\approx10-30$\,mK. Theoretical Langevin rate coefficients for the two channels are indicated by solid orange and blue lines. For reactions with N$^+_2$ in~(a), the Langevin predictions are $k^{\text{L}}_p(\text{N}_2)$=\SI{6.1e-9}{\cubic\centi\meter\per\second} and $k^{\text{L}}_s(\text{N}_2)$=\SI{3.5e-9}{\cubic\centi\meter\per\second} in the excited and ground channels, respectively. For O$^+_2$ in~(b), the corresponding values are $k^{\text{L}}_p(\text{O}_2)$=\SI{5.8e-9}{\cubic\centi\meter\per\second} and $k^{\text{L}}_s(\text{O}_2)$=\SI{3.3e-9}{\cubic\centi\meter\per\second}.  The dash-dotted lines in Figs.~\ref{fig:shut} (a) and (b) represent fits of the data to a constant function corresponding to the expected collision-energy dependence of the rate coefficient for Langevin-type dynamics. The fits yield rate coefficients $k_s(\mathrm{N}_2)$=\SI{1.5(2)e-9}{\cubic\centi\meter\per\second} and $k_s(\mathrm{O}_2)$=\SI{3.4(1)e-9}{\cubic\centi\meter\per\second} for reactions of N$^+_2$ and O$^+_2$ with Rb$(^2S_{1/2})$, respectively. In line with the results obtained in the static mode of the experiment, the rate coefficient for O$_2^+$\,+\,Rb$(^2S_{1/2})$ is in good agreement with the Langevin-capture prediction over the entire collision-energy range studied. This suggests that the dynamics is indeed dominated by long-range ion-induced dipole interactions and that the short-range reaction probability is close to unity in this channel. Conversely, the rate coefficient for N$_2^+$\,+\,Rb$(^2S_{1/2})$ is only about 1/3 of the universal Langevin value, indicating a pronounced effect of short-range interactions on the kinetics.
\noindent For N$_2^+$\,+\,Rb$(^2P_{3/2})$, both the magnitude and collision-energy dependence of the rate coefficient are consistent with a CT dominated by long-range ion-quadrupole capture in line with previous findings~\cite{hall12a}. The orange dash-dotted line in Fig.~\ref{fig:shut}~(a) represents a fit of the data for N$_2^+$\,+\,Rb$(^2P_{3/2})$ to a classical capture model including ion-quadrupole interactions yielding a quadrupole moment of $Q=15.45(1.22)$\,a.u. for the Rb~$(^2P_{3/2})$ state. This result can be compared with a theoretical value of $Q=12.9$\,a.u. computed at the MRCISD level of theory. At a collision energy $E_{\text{coll}}/k_{\text{B}}$=\SI{20}{\milli\kelvin}, a rate coefficient $k_p(\mathrm{N}_2)$=\SI{1.1(1)e-8}{\cubic\centi\meter\per\second} was determined from these data. 
\noindent For O$_2^+$\,+\,Rb$(^2P_{3/2})$ in Fig.~\ref{fig:shut}~(b), the experimental rate coefficients (orange triangles) appear to be considerably smaller than their predicted ion-quadrupole capture limit (dash-dotted orange line) and even slightly smaller than the corresponding Langevin limit (solid orange line). This finding points to a pronounced influence of short-range interactions on the dynamics, in contrast to N$_2^+$\,+\,Rb~$(^2P_{3/2})$ where this channel seems to be clearly dominated by universal long-range capture behaviour.
\subsection{Potential energy surfaces}
\label{subsec:pes}
\begin{figure}[htbp!]
	\centering
		\includegraphics[width=0.95\textwidth]{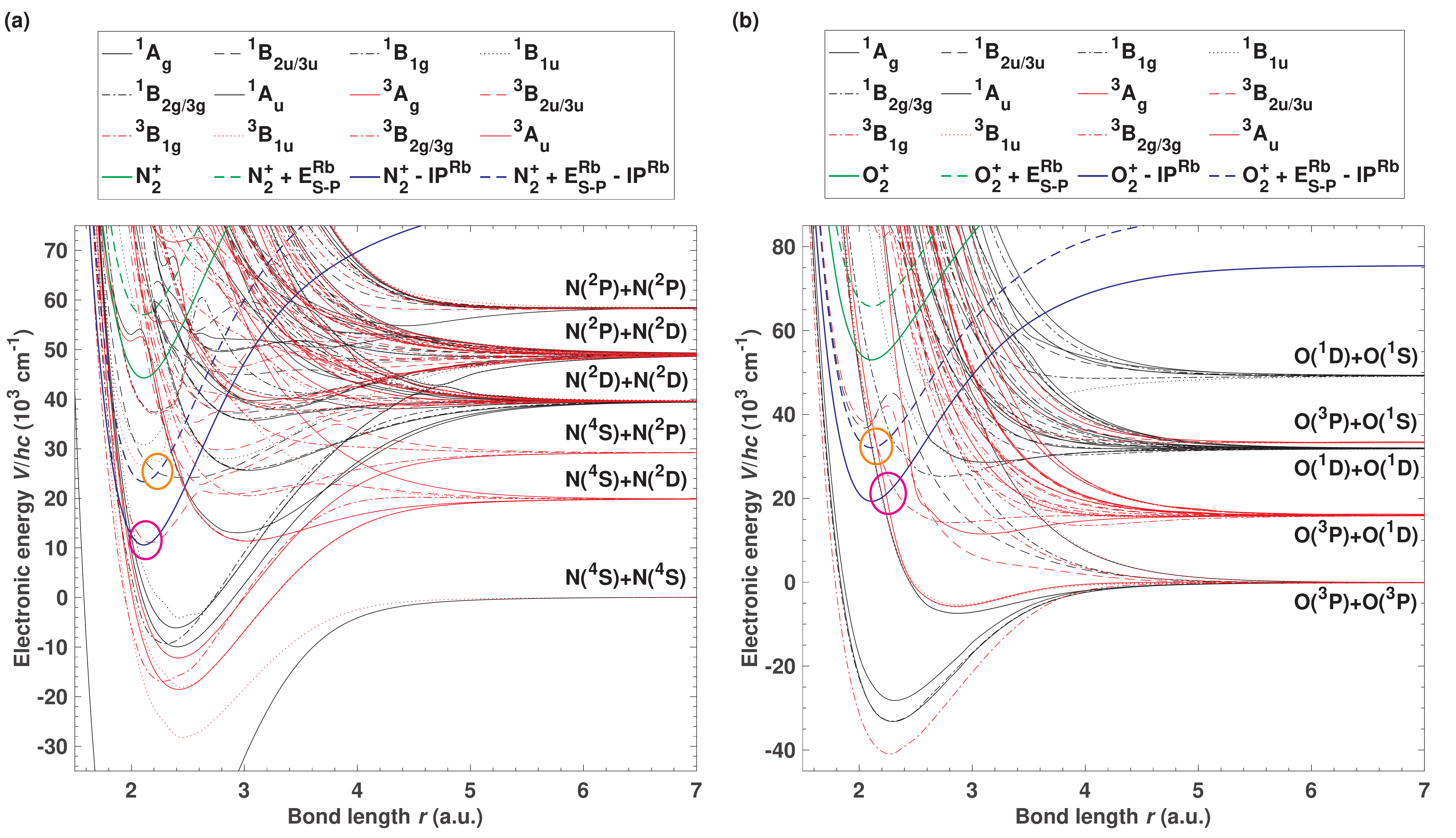}%{shuttling MOT reactions}
		\caption{Potential energy curves for (a) N$_2$ and (b) O$_2$ and their respective ions as a function of the bond length {\it r}. The states have been classified within $D_{2\text{h}}$ symmetry. The curves for the relevant cationic ground states are indicated by solid green lines, the cationic curves shifted by the ionisation potential of Rb by the solid blue lines. Dashed lines represent the corresponding curves offset by the excitation energy to the Rb $(^2P_{3/2})$ state. Circles indicate relevant crossing points promoting non-adiabatic transitions between the surfaces. See text for discussion.}
	\label{fig:pes}
\end{figure}
\noindent First insights into the widely varying CT dynamics across the systems and channels studied can be gained from analysing the relevant potential energy surfaces (PESs). Fig.~\ref{fig:pes} shows the potential curves of both singlet and triplet electronic states of the N$_2$ and O$_2$ molecules as a function of the molecular bond length $r$. The curves for the relevant cationic ground states are the solid green lines. Solid blue lines indicate the cationic curves shifted by the ionisation potential of Rb which correspond to the relevant energies of the lowest entrance channels for the CT reaction. The dashed lines represent the same curves offset by the energy of excitation to the Rb $(^2P_{3/2})$ state. The potential curves shown are cuts through the three-dimensional PES of the collision system with the Rb moiety at an infinitely large distance from the molecule. Curve crossings (indicated by circles) between the shifted ionic and neutral curves appear close to the molecular equilibrium geometry in both systems providing opportunities for non-adiabatic transitions and therefore CT around the crossing points.
\begin{figure}[b!]
	\centering
		\includegraphics[width=1\columnwidth]{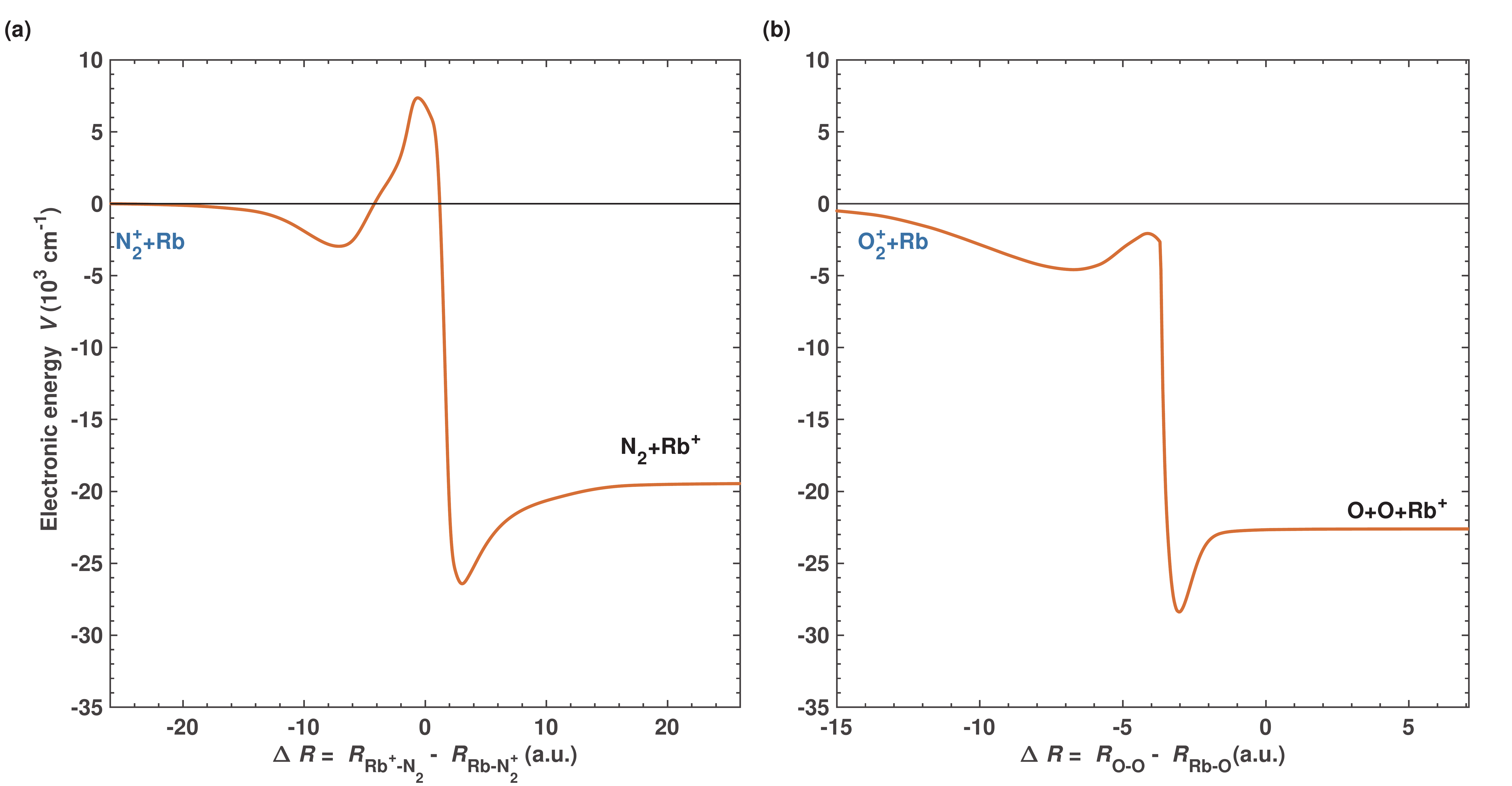}%{shuttling MOT reactions}
		\caption{Cuts of the PES along reaction coordinates for CT in the linear collision geometry of the singlet channels of Rb($^2S_{1/2}$) reacting with \textbf{(a)}~N$_2^+$ and \textbf{(b)}~O$_2^+$. See text for details.}
	\label{fig:rxncoord}
\end{figure}
\noindent As the collision partners all have doublet electron-spin character, collisions can occur on either singlet or triplet potential energy surfaces. Fig.~\ref{fig:rxncoord} shows cuts of the adiabatic potential surfaces for (a) N$_2^+$\,+\,Rb~$(^2S_{1/2})$ and (b) O$_2^+$\,+\,Rb~$(^2S_{1/2})$ close to the computed minimum energy path for CT in the singlet channels at a linear collision geometry. For N$_2^+$\,+\,Rb, an electronic barrier is found along this reaction path. While a CT reaction along this coordinate is thus in principle possible, the height of the barrier exceeds 5000\,cm$^{-1}$ and thus prevents CT at the low collision energies of order $\approx 10^{-2}$\,cm$^{-1}$ of the present experiments. The potential-energy profile for O$_2^+$\,+\,Rb$(^2S_{1/2})$ in Fig.~\ref{fig:rxncoord}~(b) also exhibits a barrier along the shown reaction coordinate for CT. In this case, however, the barrier is submerged and thus provides no significant impediment to CT at the experimental collision energies. A similar curve crossing can also be found in the corresponding triplet reaction channel (see the red curve in Fig.~\ref{fig:pes}~(b) crossing the entrance channel close to the equilibrium geometry within the magenta circle). Thus, together with the experimental findings it can be surmised that the short-range CT probability is near unity in both the singlet and triplet collision channels for O$_2^+$+Rb$(^2S_{1/2})$ and that the CT rate coefficient is near the capture limit dominated by long-range Langevin-type interactions.
\begin{figure}[htbp!]
	\centering
		\includegraphics[width=1\columnwidth]{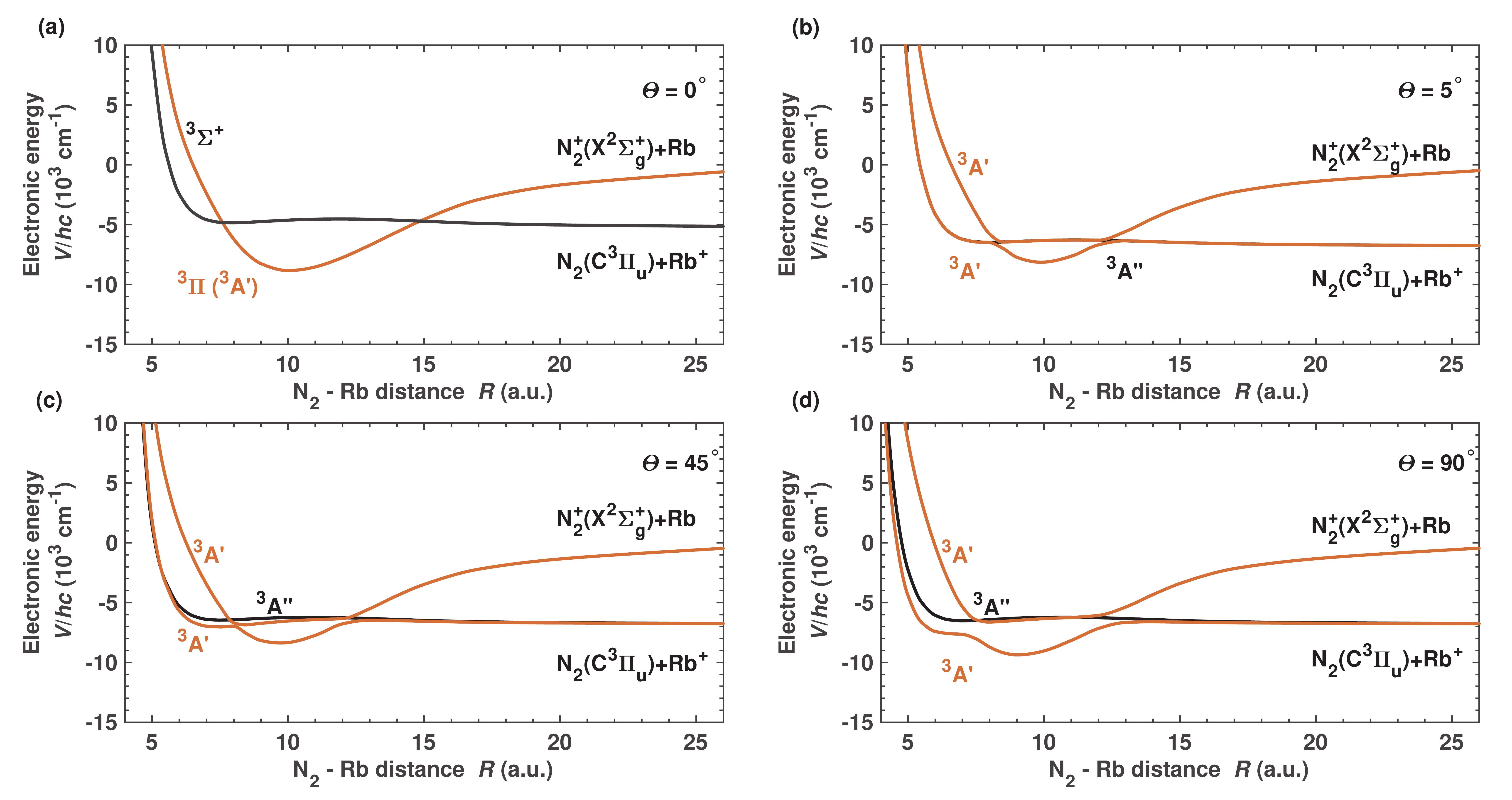}
		\caption{Cuts of the PES in the triplet channel for CT reactions of Rb($^2S_{1/2}$) with N$_2^+$ at different orientation angles $\theta$. $R$ denotes the N$_2$\,-\,Rb separation.}
	\label{fig:pescut}
\end{figure}
\noindent Since CT in N$_2^+$\,+\,Rb~$(^2S_{1/2})$ is unlikely to occur via the singlet collision channel, Fig.~\ref{fig:pescut} shows cuts of the triplet PES of the collision systems along the N$_2$-Rb coordinate $R$ for different N$_2$-Rb orientation angles $\theta$. In the linear geometry ($\theta=0$, Fig.~\ref{fig:pescut}~(a)) the entrance channel with symmetry $^3\Sigma^+$ crosses with a $^3\Pi$ surface asymptotically connecting to the $C~^3\Pi_u$ state of N$_2$. In a non-linear ($\theta \neq 0$) collision geometry (Figs.~\ref{fig:pescut}~(b)-(d)), the $^3\Pi$ surface splits into an $A'$ and an $A''$ component with the $A'$ surface undergoing avoided crossings with the entrance channel surface (also $A'$ in this symmetry). The separation of the resulting adiabatic surfaces at the crossing points, and hence the non-adiabatic coupling between them, increases with increasing orientation angle $\theta$. The probability of non-adiabatic transitions around the crossings, and therefore CT, is thus expected to be strongly dependent on the orientation of the collision partners. 
\subsection{Quasiclassical trajectory simulations and quantum dynamics calculations}
\noindent To gain quantitative insight into the non-adiabatic dynamics of N$_2^+$\,+\,Rb~$(^2S_{1/2})$, we performed both quasiclassical trajectory simulations and quantum scattering calculations of the CT. The classical trajectory simulations employed a modified Landau-Zener formalism on the two coupled two-dimensional $A'$ surfaces of Fig. \ref{fig:pescut} for modeling the CT. In the simulations, the bond length was frozen at the equilibrium value for N$_2^+$, thus yielding a two-dimensional dynamics along the distance $R$ between the centres-of-mass of N$_2^+$ and Rb and the orientation angle $\theta$. Rate coefficients have been calculated for $E_\text{coll}/k_{\rm B} = 10$, 20 and 30\,mK from running 5000 independent trajectories each. The CT rate coefficients obtained are shown as red diamonds in Fig.~\ref{fig:shut} and are in good agreement with the experimental values. Further insights into the CT dynamics can be gained from an analysis of the trajectories. Reactive CT trajectories are divided into two categories: (i) trajectories with a single collision, i.e., direct trajectories, and (ii) trajectories with multiple collisions, i.e., indirect trajectories. To categorise the trajectories, we classify them according to the collision time defined as the time elapsed between the first and last time a trajectory satisfies a geometrical criterion, here the sum of the three inter-atomic distances has to be smaller than 35\,a.u.. Most of the trajectories have collision times around $\sim 0.5$\,ps and the processes are, therefore, direct. At $E_\text{coll}/k_\text{B}=20$\,mK, 4374 of the 5000 trajectories show charge transfer (3294 direct and 1080 indirect) whereas 626 end as N$_2^+$+Rb (with 67 flyby or no collision, 513 direct and 46 indirect).
\begin{figure}
\includegraphics[width=0.95\columnwidth]{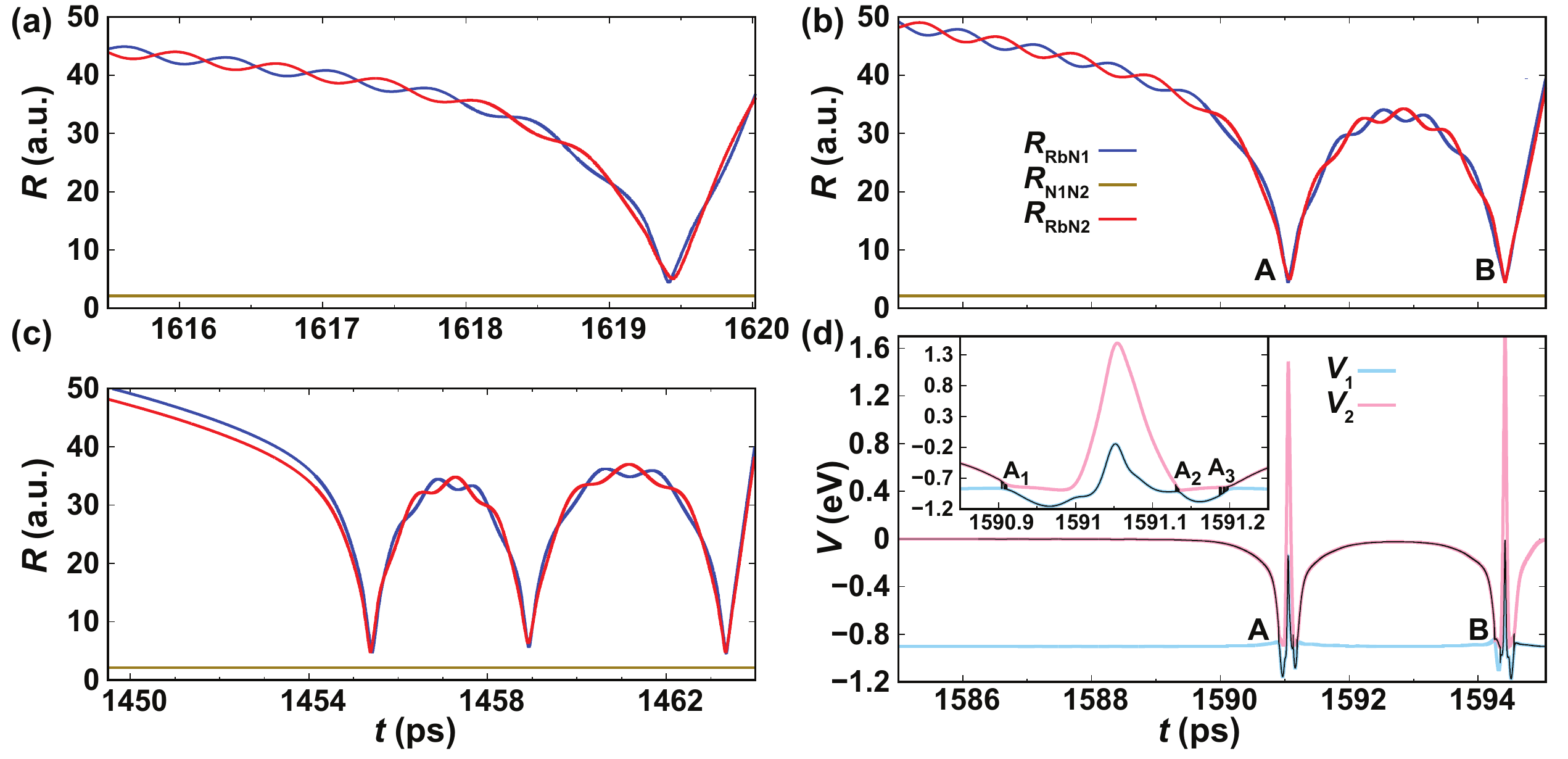}
\caption{Direct and indirect CT trajectories. Trajectories are shown in terms of the N$_2$-Rb distance $R$ as a function of simulation time $t$. Panel (a) represents a direct trajectory while panels (b) and (c) show two indirect trajectories with multiple collisions. The trajectory in panel (b) is further analysed in panel (d) which shows the change in potential energy with respect to $t$. The potential energies $V_1$ and $V_2$ of the coupled electronic states are shown as faint blue and red lines, respectively. The black line is the energy path of the trajectory. Three hopping regions for the first collision (labeled $A_1$ to $A_3$) can be seen in the inset. Multiple switching between the surfaces are observed in the hopping regions.}
\label{fig:ct1}
\end{figure}
\noindent Three illustrative example trajectories, one direct and two indirect, are shown in Fig. \ref{fig:ct1}. The dynamical path of the indirect trajectory with two collisions shown in Fig. \ref{fig:ct1} (b) is displayed in Fig.~\ref{fig:ct2} as its projection onto the two PESs. Large parts of the available configurational space are sampled despite the low collision energy. Crossings of the trajectory between the two PESs are distributed along all values of $\theta$ and concentrated around $R \sim 8$\,a.u. and 12\,a.u. as can be expected from the 1D cuts through the PES shown in Fig.~\ref{fig:pescut}. Multiple recrossings (labelled as events ``A'' and ``B'') occur over the duration of the dynamics (indicated by the changing colours of the trajectory).
\begin{figure}
\includegraphics[width=0.95\columnwidth]{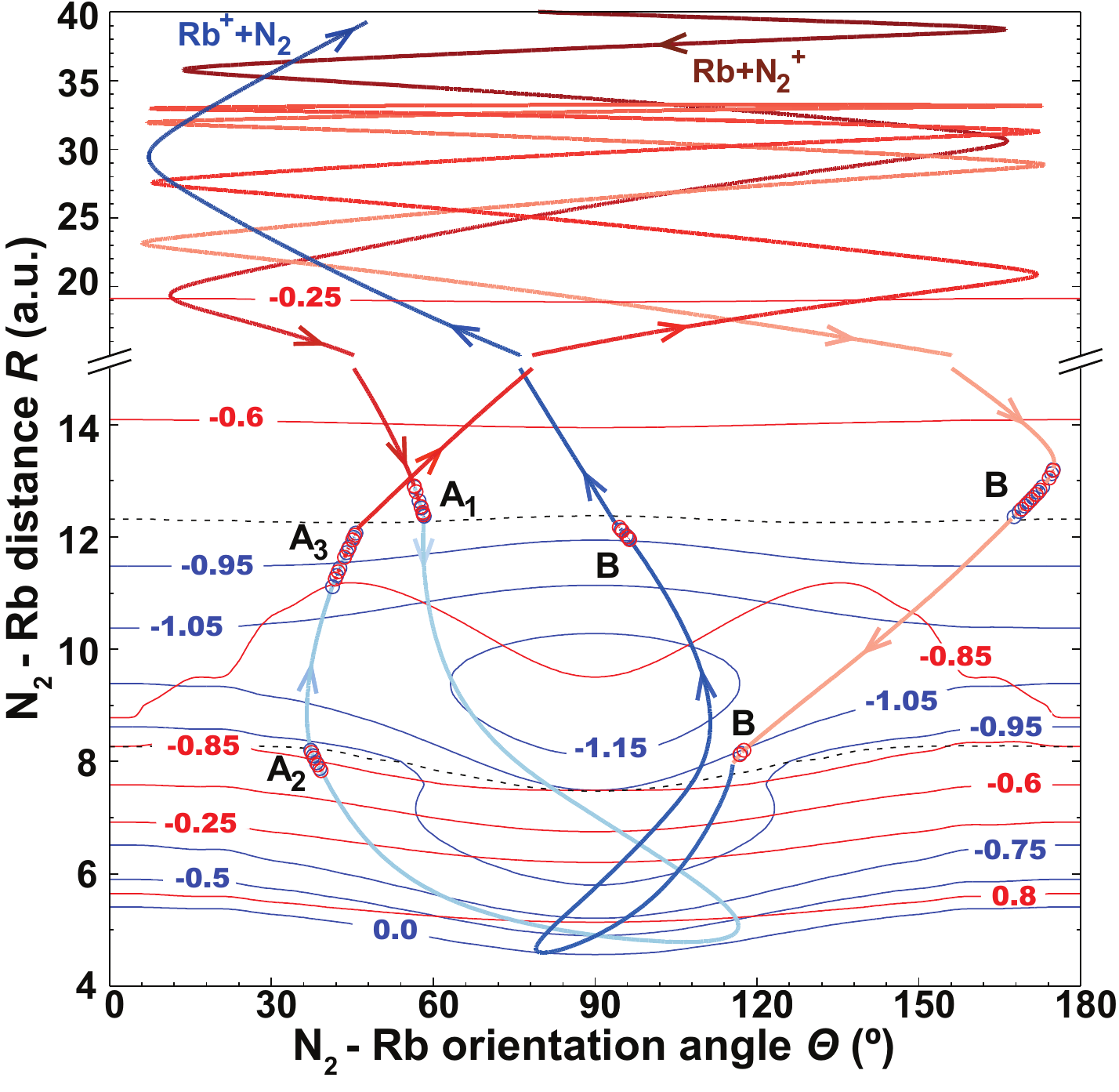}
\caption{ Detailed dynamics of the trajectory shown in Fig.~\ref{fig:ct1}~(b) and (d) represented as a projection onto the two coupled PESs. Contour diagrams of the two PESs are shown as solid red (upper state) and blue (lower state) lines. Contour lines are labeled in eV. The avoided crossing regions are shown as black dashed lines. Thick red and blue lines show the trajectory path on the upper and lower surface, respectively. The progress in time of the trajectory is represented as a change in colour from dark to light (for the upper state) and light to dark (for the lower state) colour tones. Hops between surfaces are indicated as red and blue open circles in regions labelled identically to \ref{fig:ct1}~(b) and (d). See text for discussion.}
\label{fig:ct2}
\end{figure}
\noindent The trajectories, both direct and indirect, typically show multiple crossings between the surfaces (note that even a direct trajectory typically traverses four crossing regions). After the first collision, an outward trajectory ending up on the upper (entrance) surface is frequently trapped in the deep potential well as the low initial collision energy has been redistributed into other internal degrees of freedom and is not available for dissociation of the complex anymore. These trajectories often show pronounced large-amplitude internal rotations of the complex, i.e., the Rb orbits the N$_2$ moiety at large distances several times, see Fig.~\ref{fig:ct2}. These trajectories can then undergo multiple collisions. However, once an outward trajectory ends up on the lower (CT) surface after traversing the outer crossing point, it cannot return anymore because of the repulsive character of the lower potential surface in this region (see Fig.~\ref{fig:pescut}). Thus, the distinct short-range topology of the two coupled surfaces ultimately contributes to explaining the high, but not unit, efficiency for CT in this channel. It is also evident that a long-range Langevin-capture picture cannot explain the complex short-range dynamics and its impact on the kinetics as observed in the experiments. The insights from classical dynamics were corroborated by 1D quantum-scattering calculations of the CT performed at different orientation angles $\theta$. This 1D treatment is in the spirit of an infinite-order sudden approximation (IOSA) previosuly employed for ion-neutral collisions~\cite{tsien73a}. The quantum results for the CT rate coefficients averaged over all orientation angles are indicated by the green symbols in Fig.~\ref{fig:shut}~(a) and are in good agreement with both the experimental and classical-dynamical results. From the good agreement between experimental results and the two different theoretical treatments of the CT rate coefficients for N$_2^+$\,+\,Rb~$(^2S_{1/2})$, several conclusions can be drawn. First, the CT dynamics can be understood in classical terms and apart from surface hopping, distinct quantum effects such as tunneling and zero-point motion do not seem to play a major role due to the large mass of the collision partners involved. Second, it appears adequate to approximate the dynamics in reduced dimensionality, i.e., in 2D or even in 1D within an IOSA-type approximation, in order to correctly reproduce the observed kinetics. Note, however, that this cannot be expected for O$_2^+$\,+\,Rb~$(^2S_{1/2})$ in which case the reaction coordinate is more complex, see Fig.~\ref{fig:rxncoord}~(b). Note also that an 1D treatment cannot capture the roaming-type behaviour which seems to be important for certain types of collisions. Third, in spite of the complex and dense electronic structure in the energy region of the entrance and exit channels (see Fig.~\ref{fig:pes}), it seems sufficient to describe CT in N$_2^+$\,+\,Rb~$(^2S_{1/2})$ by including only two coupled PESs. This can be rationalised in terms of the strong non-adiabatic couplings which happen around the crossing points between the two surfaces. Couplings to other states which show no crossings with the entrance channel in the energy region sampled by the experiments are expected to be considerably weaker and do not appear to affect the CT dynamics appreciably. Fourth, the QCT calculations reveal that the reaction mechanism is a combination of direct and complex-forming collisions. The latter typically show large-amplitude orbiting motions of the Rb around the N$_2$ moiety  before the reaction occurs. This behaviour is somewhat reminiscent of the roaming dynamics recently discovered in a range of polyatomic reaction systems~\cite{townsend04a, joallande14a, bowman17a}. Indeed, the type of dynamics uncovered here can be expected to be a common feature in cold reactions which proceed via the formation of a reactive complex with very little excess energy~\cite{tong12a}. Note that in a full 3D model of the dynamics, one can also expect the excitation of the N-N stretching vibration in the complex which was frozen in the present 2D treatment potentially leading to an even more diverse collision dynamics. A quantitative modelling of the dynamics of the O$_2^+$+Rb~$(^2P_{3/2})$ CT was outside the scope of the present study because of the considerably increased computational cost of the quantum-chemistry calculations for this excited channel. 
\section{Conclusions}
\label{sec: Conclusion}
\noindent The cold CT dynamics explored in the present molecular collision systems has to be contrasted with the results of the atomic systems studied so far. In the vast majority of cases reported, CT was observed to be slow, i.e., the rate coefficients were found to be several orders of magnitude smaller than the universal capture limit, and dominated by radiative couplings~\cite{hall13a,zipkes10b,schmid10a,sikorsky18a,joger17a,haze15a,dasilva15a}. Notable exceptions are specific channels in Ca$^+$\,+\,Rb~\cite{hall11a}, Yb$^+$\,+\,Ca~\cite{rellergert11a}, Yb$^+$\,+\,Rb~\cite{ratschbacher12a}, Yb$^+$\,+\,Li~\cite{joger17a} and Ca$^+$\,+\,Li~\cite{haze15a} in which CT was found to be non-adiabatic, but still considerably slower than the capture limit. By contrast, the CT rate coefficients of the systems considered here were all found to be close to the capture limit or only slightly slower. The good agreement between the experimental and the theoretical non-adabatic CT rate coefficients suggest that CT is indeed dominated by non-adiabatic effects in the present case. This conclusion is also corroborated by a recent theoretical study which found that radiative couplings are small in the N$_2^+$\,+\,Rb system~\cite{gianturco19a}. Because the increased complexity of molecular compared to atomic collision systems provides more numerous opportunities for channel crossings, it can be expected that the situation observed here for N$_2^+$\,+\,Rb and O$_2^+$\,+\,Rb is fairly general. It can be surmised that CT will often be nonadiabatic and fast in the molecular systems of interest for cold-collision studies. 
\noindent The trends observed here could be rationalised in terms of the efficiency of the nonadiabatic couplings involved. When the nonadiabatic transition probability during a collision is close to unity, the kinetics can be modelled by universal classical capture theory and is governed by the specific long-range interactions in the system, as observed here for N$_2^+$\,+\,Rb$(^2P_{3/2})$ and O$_2^+$\,+\,Rb$(^2S_{1/2})$. Otherwise, the exact CT rates depend on the specific positions of curve crossings and the strengths of the relevant nonadiabatic couplings as well as on the topologies of the PESs involved, as observed here in N$_2^+$\,+\,Rb$(^2S_{1/2})$. The short-range dynamics in this system was found to exhibit multiple transits of crossing regions in single and multiple collision events with the latter showing large-amplitude internal motions of the reaction complex. In this case, the effects of the long-range dynamics on the kinetics are superseded by short-range effects which cannot be predicted without detailed theoretical modelling. This illustrates that it cannot, a priori, be anticipated whether universal behaviour applies and in particular whether the universal Langevin picture which is often invoked in the explanation of cold ion-neutral reactive processes is valid. 
\section{Acknowledgements}
\noindent The current work was supported by the Swiss National Science Foundation, grant nr. 200020\_175533, the NCCR MUST and the University of Basel. M.T.~was supported by the National Science Centre Poland (Sonata Grant nr.~2015/19/D/ST4/02173) and the PL-Grid Infrastructure.
\section{Author contributions}
\noindent A.D. performed all experiments except for the reactions of O$_2^+$ with a stationary Rb cloud which were performed by P.E. S.W. supervised the experimental part of the project. M.T. performed the electronic-structure and quantum-dynamics calculations. D.K. and M.M. performed the classical-dynamics calculations. A.D., M.T., D.K., M.M. and S.W. wrote the manuscript.
\bibliography{Main-Dec18,newref}% Produces the bibliography via BibTeX.
\end{document}